\documentclass[final,5p,times,twocolumn]{elsarticle}

\usepackage{graphics}
 \usepackage{graphicx}
 \usepackage{epstopdf}
 \usepackage{epsfig}
\usepackage{amssymb}
\usepackage{amsmath}
\usepackage{lineno}

\usepackage{tikz}
\usepackage{lipsum}
\usepackage{subfig}
\usepackage{booktabs}

\begin{document}
\begin{frontmatter}
\title{An ultracold electron facility in Manchester}
\author{O. Mete\corref{cor1}\fnref{fn1}}
\ead{oznur.mete@manchester.ac.uk}
\author{R. Appleby\fnref{fn1}} 
\author{W. Bertsche\fnref{fn2}}
\author{S. Chattopadhyay\fnref{fn1,fn3}}
\author{M. Harvey\fnref{fn4}}
\author{A. Murray\fnref{fn4}}
\author{G. Xia\fnref{fn1}}
\address[rvt]{The University of Manchester, Manchester, M13 9PL, United Kingdom} 
\cortext[cor1]{Corresponding author.}
\fntext[fn1]{The Cockcroft Institute of Accelerator Science and Technology.}
\fntext[fn2]{The European Organisation for Nuclear Research, CERN.}
\fntext[fn3]{The University of Liverpool and The University of Lancaster.}
\fntext[fn4]{The University of Manchester - Photon Science Institute.}
\begin{abstract}
An ultra-cold atom based electron source (CAES) facility has been built in the Photon Science Institute (PSI), University of Manchester. In this paper, the key components and working principles of this source are introduced. Pre-commissioning status of this facility and preliminary simulation results are presented.
\end{abstract}
\end{frontmatter}
\section{Introduction}
Imaging of ultrafast biological or chemical processes is often achieved using X-rays or electrons \cite{Henderson, Engelen}. Both methods possess certain advantages for different cases. X-ray free electron lasers (XFELs) can be used to provide the required atomic resolution for diffraction experiments whereas a table-top low energy electron beam could provide orders of magnitude higher resolution at a significantly lower cost. Also the elastic scattering cross-section of electrons is larger than photons, so that the required intensity is lower when electrons are used. On the other hand, the penetration depth is larger for X-rays, therefore probing with them is more advantageous to study thicker samples whereas electrons are applicable to study surfaces, thin films and membrane proteins. In addition, lower energy deposition through elastic and inelastic scattering of electrons in biological specimens provides a less destructive diagnostics method. Classical electron microscopes produce electrons from a cathode-tip and can provide temporal and spatial resolutions in the order of ps and nm, respectively, by using various techniques \cite{eMicr-ps, eMicr-nm}. Consequently, in order to reach higher resolutions needed for ultrafast phenomena, other methods which can produce a beam of about $10^8$ electrons are essential. 
\begin{figure}[htb!]
\centering
\includegraphics[width=0.50\textwidth] {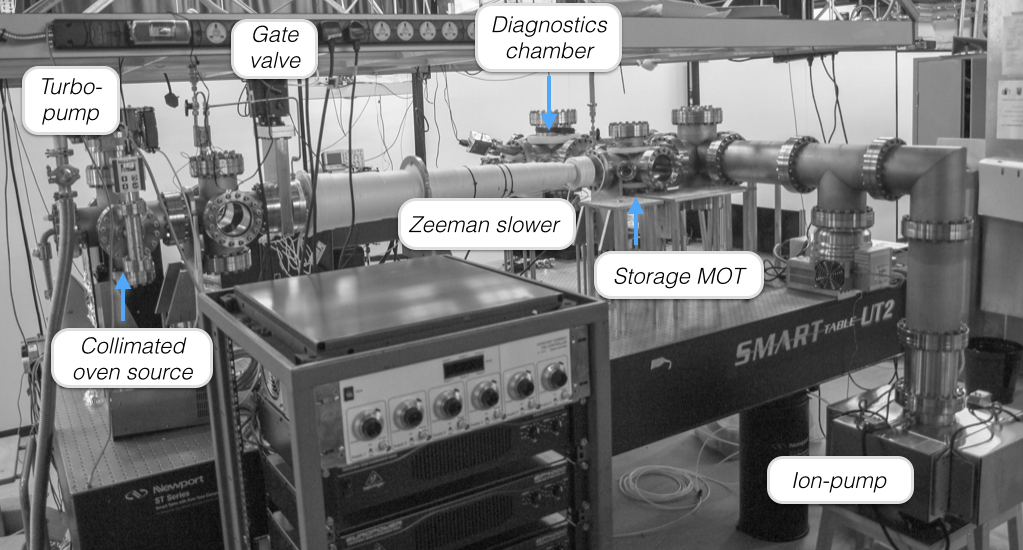}
\caption{Layout of the facility.}
\label{fig:layout}
\vspace{-0.5em}
\end{figure}
In this context, probing electrons can be produced by a photo-injector or a cold electron source depending on the intensity, emittance and energy spread required by the experiment. Photo-injectors are high intensity low emittance electron sources where space charge force plays a significant role which might lead to blurred images when used for microscopy. Instead the ultra-cold atom trap based electron sources are excellent candidates where a moderate intensity electron beam can be produced with low emittance for diffraction studies \cite{Taban} and low energy spread for spectroscopy (where absorption and scattering cross-sections are critically dependent on energy) according to the extraction method used. 

\begin{figure}[htb!] 
\includegraphics[width=0.48\textwidth] {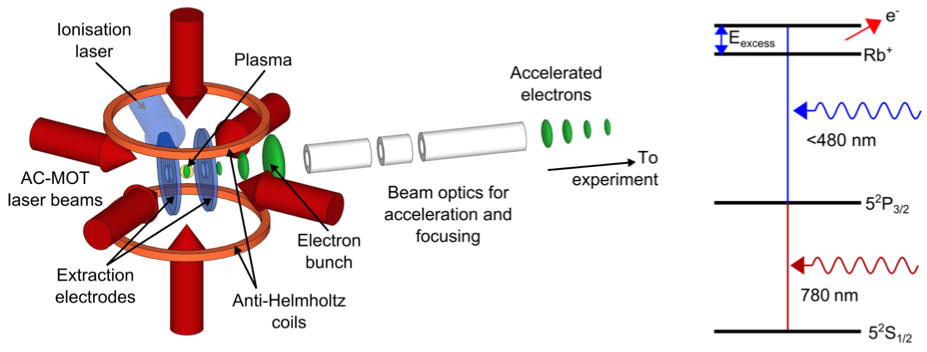}
\caption{Magneto-optical atom trap configuration (left) and the photo-ionisation process (right).}
\label{fig:trap}
\vspace{-1.5em}
\end{figure}

Manchester's ultra-cold atom trap based electron source setup consists of an atom source, Zeeman slower, magneto-optical trap, storage and diagnostics chambers as presented in Fig.\ref{fig:layout}. An atomic beam is produced using a Rubidium (Rb) oven. The slowing of the atoms is performed through deceleration with a counter propagating laser beam by momentum transfer from the laser field. The fundamental problem of the scheme is Doppler shift of atoms out of resonance with respect the laser beam. This originates partly from the certain line width of the laser and partly from the initial momentum distribution of the atoms. Doppler shift can be compensated by using a spatially varying magnetic field for shifting resonance atomic transition to match the laser frequency. This is provided by a setup called Zeeman Slower that consists of a solenoid with varying magnetic field. Confinement of the atoms are achieved in a magneto optical trap (MOT). Three pairs of circularly polarised, counter propagating orthogonal laser beams and an inhomogeneous magnetic field provided by anti-Helmholtz coils form the atom trap, as shown in Fig. \ref{fig:trap}. The trapping laser beams are centred on the quadrupole field of the coils. An electron beam can be produced by ionising the ultra cold atom cloud confined in the MOT by using an additional laser beam (ionisation laser). A new technique, the AC-MOT, was invented in the University of Manchester \cite{Harvey}. This technique relies on the fast switching ($<20\,\mu s$) of the magnetic field on the coils and therefore provides a magnetic field free time for production and extraction of the electrons. This is important in order to prevent the electrons gaining transverse momentum due to the magnetic field which will lead to the growth of the beam emittance \cite{archive}. An atom cloud can be maintained in the storage MOT and it can be pushed to the experimental chamber by using a laser beam. For the initial commissioning, the experimental chamber contains the electron extraction electrodes and a multi-channel plate (MCP) detector for beam profile measurements. 

Laser beams will be provided from the high resolution tuneable CW lasers in the Photon Science Institute of the University of Manchester. For atom slowing and trapping, a CW Ti:Sapphire laser with the wavelength of $780.24\,nm$ will be used, whereas the ionisation will be performed with CW and pulsed laser beams of $296\,nm$ and $480\,nm$ to study ionisation from different Rb energy levels \cite{archive}.   
%
\section{Extraction of cold electrons}
Ionised electrons will be separated and transferred out of the MOT towards the diagnostic stages or onto an experiment. A set of three electrodes is used to provide the extraction field. Electrodes are made out of fine mesh that is transparent to the laser beams and still provides 99$\%$ transmission for the electrons. Bias voltages and the locations of the electrodes can be configured to provide different initial beam characteristics. Beam tracking was performed by using General Particle Tracer code (GPT) \cite{GPT}. 3D space charge force was taken into account to study the extraction process. The setup can function either in "penetrating field" or in "high voltage" mode to provide low energy spread or low emittance, respectively. 
\subsection*{Penetrating field model}
\begin{figure}[htb!]
\centering
\includegraphics[width=0.54\textwidth] {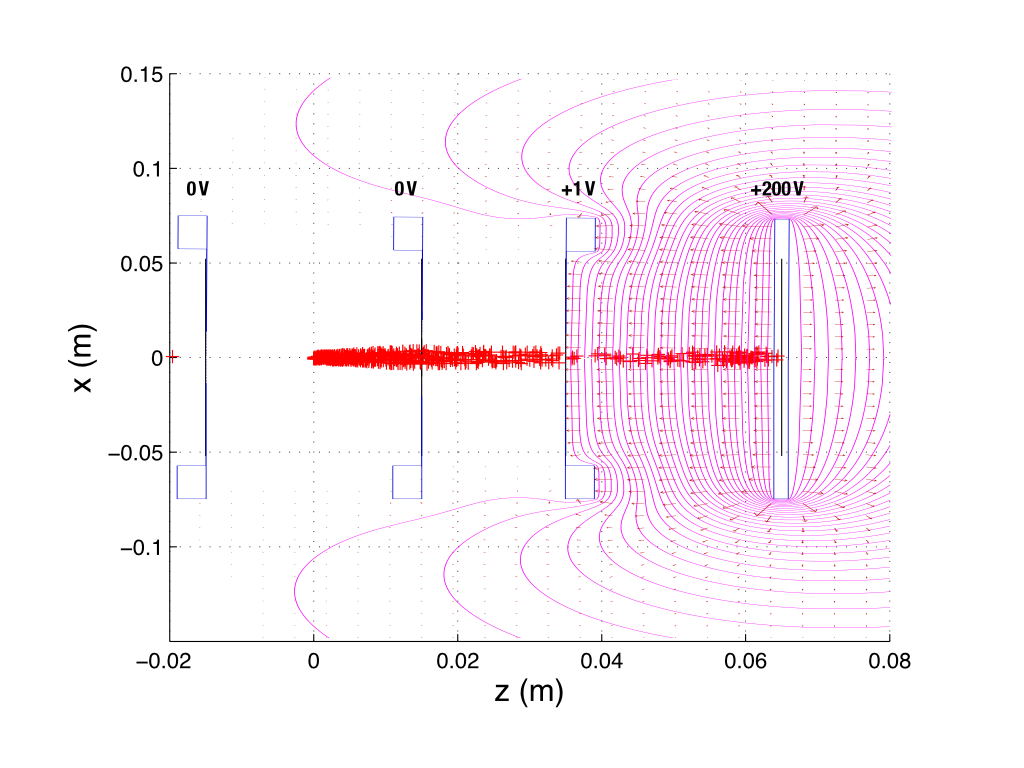}
\caption{ Electron trajectories under the effect of the given field across the electrodes and the MCP. Electrons are created at position $0$ and move from left to right.}
\label{fig:field}
\vspace{-1.5em}
\end{figure}

\begin{figure}[htpb!]  
\centering
\includegraphics[width=0.50\textwidth] {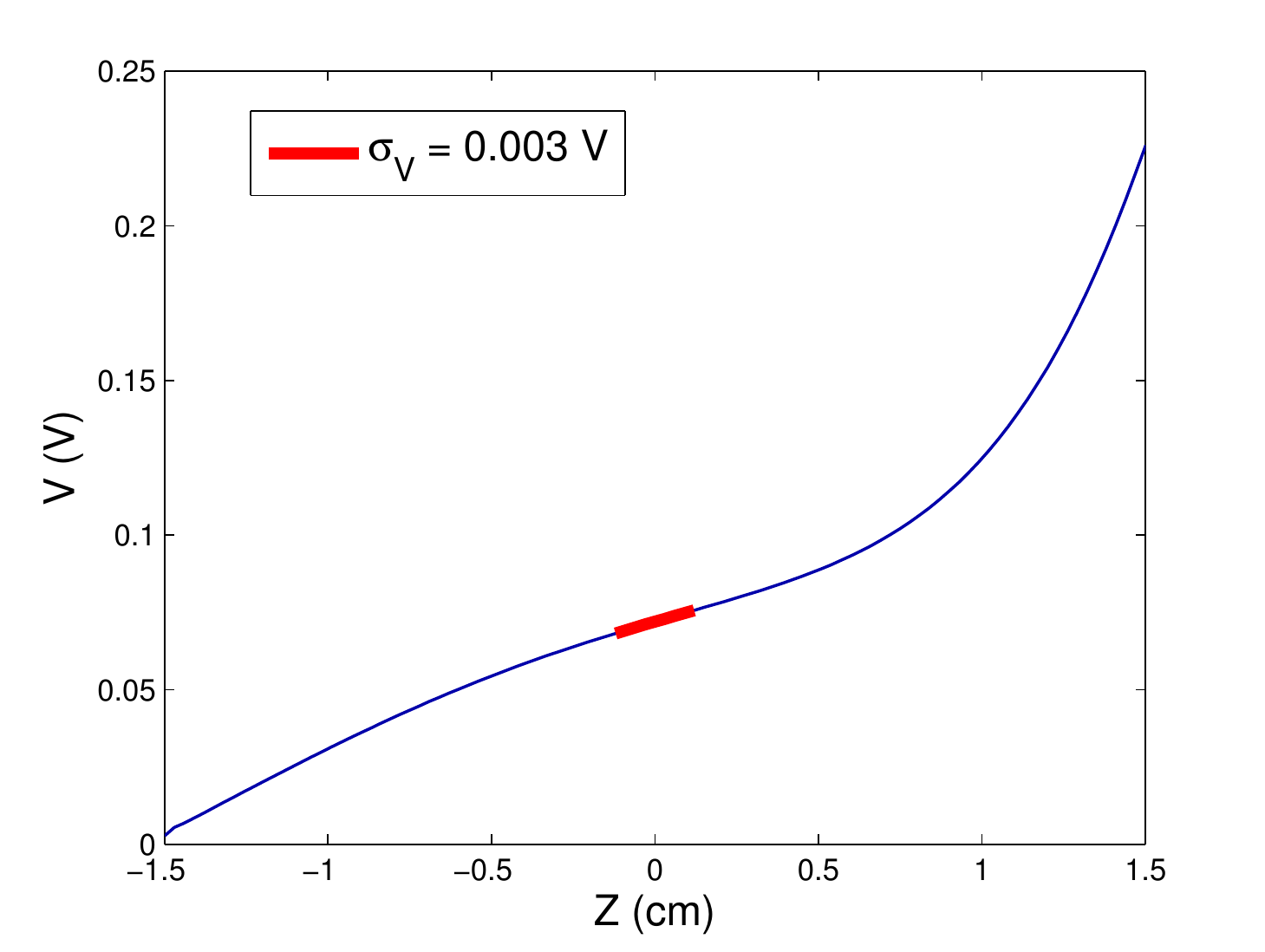}
\caption{Voltage seen by the electrons travelling in the $z$ axis. Location of the MOT is denoted by the red line.}
\label{fig:voltage_scheme2}
\vspace{-1.5em}
\end{figure}

\begin{figure}[htpb!] 
\centering
\includegraphics[width=0.50\textwidth] {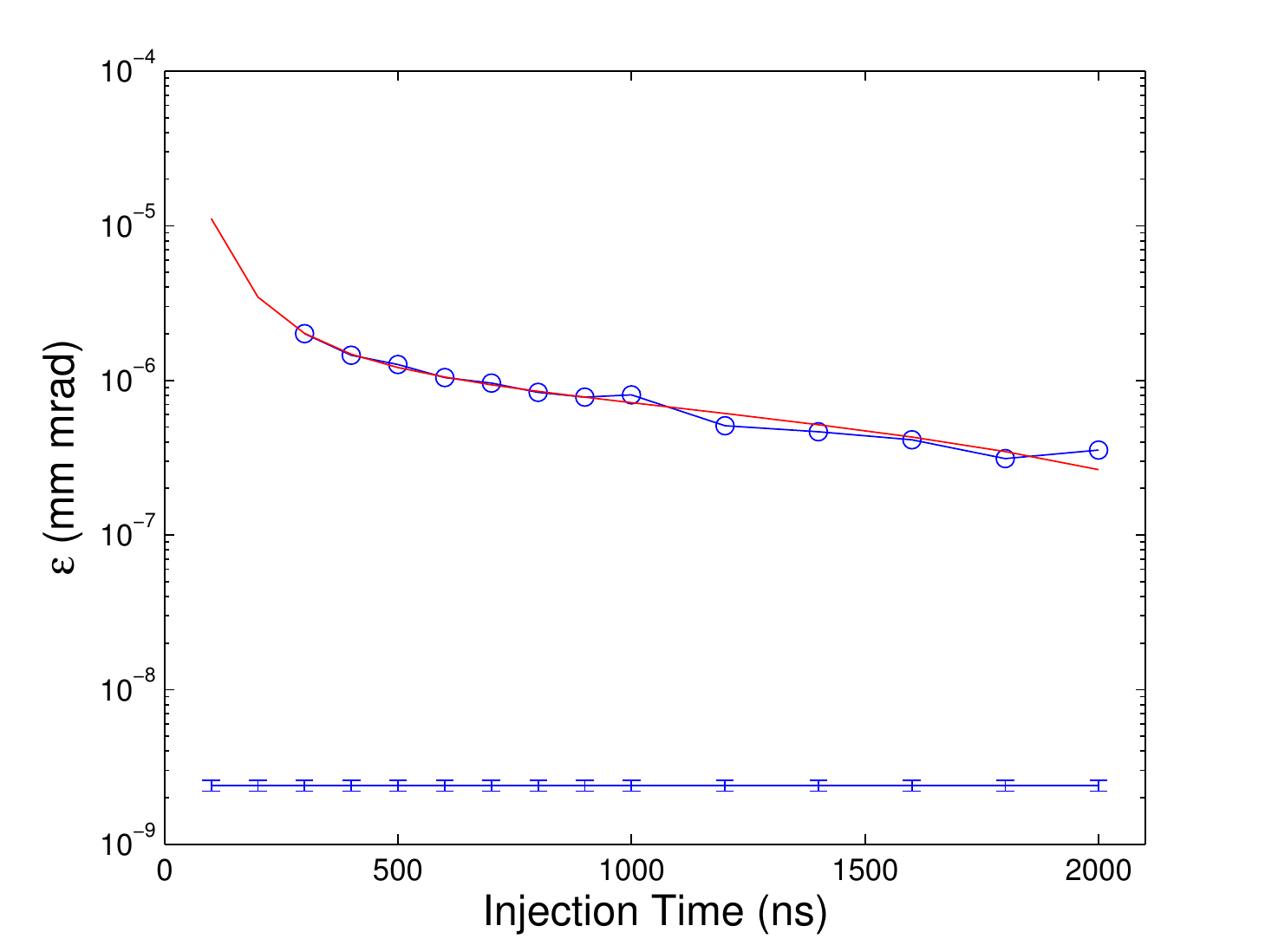}
\caption{Transverse normalised emittance at the location of the MCP, scaling with ionisation time (i.e., laser pulse length) from $100\,ns$ to $2\, \mu s$.}
\label{fig:emitt_scheme2}
\vspace{-1.5em}
\end{figure}

The initial energy spread within the extracted electrons can be minimised by using a penetrating field model. This model relies on a very low potential difference between the electrodes which will limit the energy difference between the electrons, as they are created, as shown in Fig.\ref{fig:field}. Consequently, the first two electrodes are grounded and the third one is kept at $1\,V$. Once they are separated from the ions in the MOT, the electrons are accelerated towards the front of the MCP detector that is kept at $+200\,V$. In this scheme a field of a fraction of a $V/cm$ penetrates into the MOT through the hole in the centre of the second electrode. The voltage deviation across the MOT, $\sigma_V$, is $3\,mV$ as seen in Fig.\ref{fig:voltage_scheme2}. 
The scheme allows electrons to be produced with the same initial energy provided by the laser; $300\,meV$ was considered for the results presented in this paper. However due to the lack of rapid acceleration, the space charge force is strong enough to cause a large beam emittance. This can be mitigated by ionising the same number of electrons ($1\,pC$) within a longer time scale thereby diluting the charge density. This method is feasible for the case where a need for energy resolution exceeds the need for a low emittance beam and intensity. Nevertheless, the emittance can be reduced below $1\,mm\,mrad$ for injection times longer than $600\,ns$, as demonstrated in Fig.\ref{fig:emitt_scheme2}. Beam emittance should reach a point where space charge force is negligible while the photo-ionisation time approaches to infinity, as denoted by the straight line for the space charge free case where emittance is $2.5\,nm \pm0.2nm$. 
%
\subsection*{High voltage configuration}
\begin{figure}[htb!] 
\centering
\includegraphics[width=0.50\textwidth] {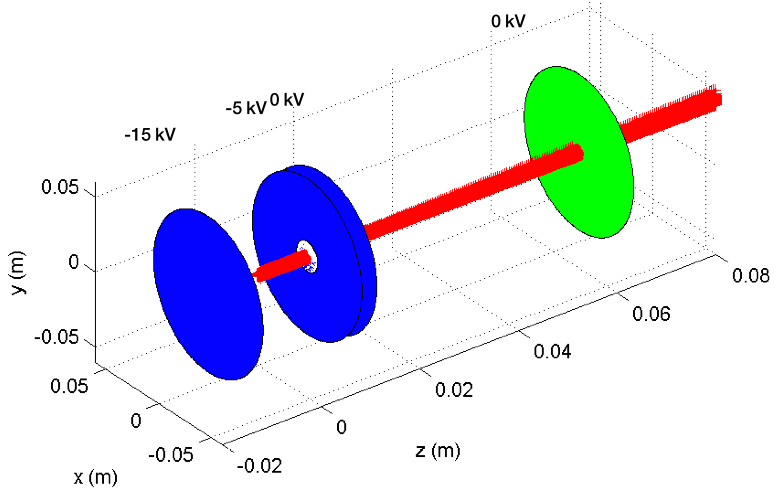}
\caption{Electron trajectories for the low emittance configuration. Blue discs are the electrodes, green disc is the MCP with voltages of $-15\,kV$, $-5\,kV$, $0\,V$ and $0\,V$, respectively.}
\label{fig:traj1}
\vspace{-2em}
\end{figure}

\begin{figure}[htpb!] 
\centering
\includegraphics[width=0.50\textwidth] {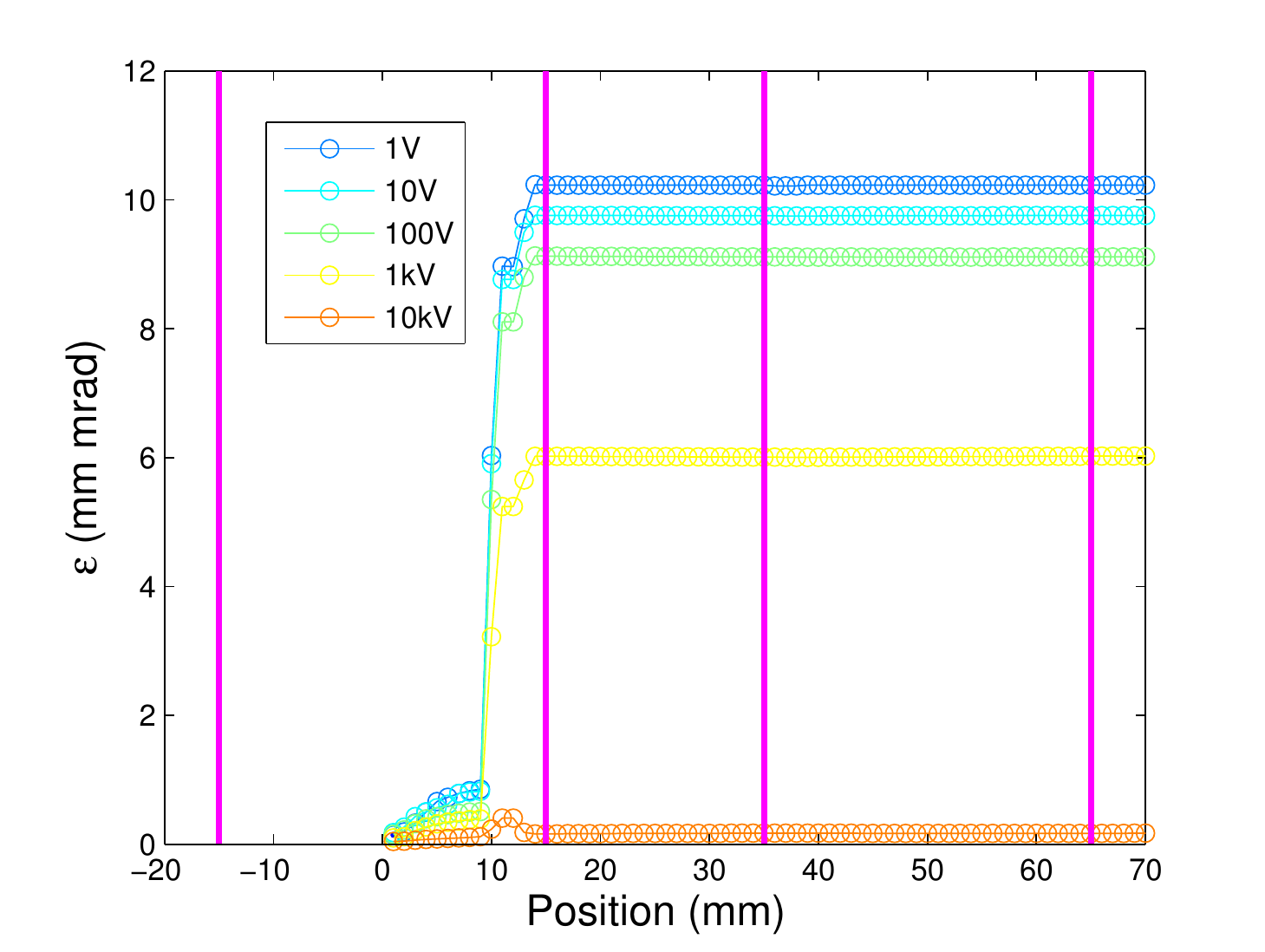} 
\caption{Evolution of space charge dominated emittance under various extraction voltages. Locations of electrodes are denoted with pink stems.}
\label{fig:emitt_scheme1}
\vspace{-1.5em}
\end{figure}

\begin{figure}[htpb!]
\centering
\includegraphics[width=0.50\textwidth] {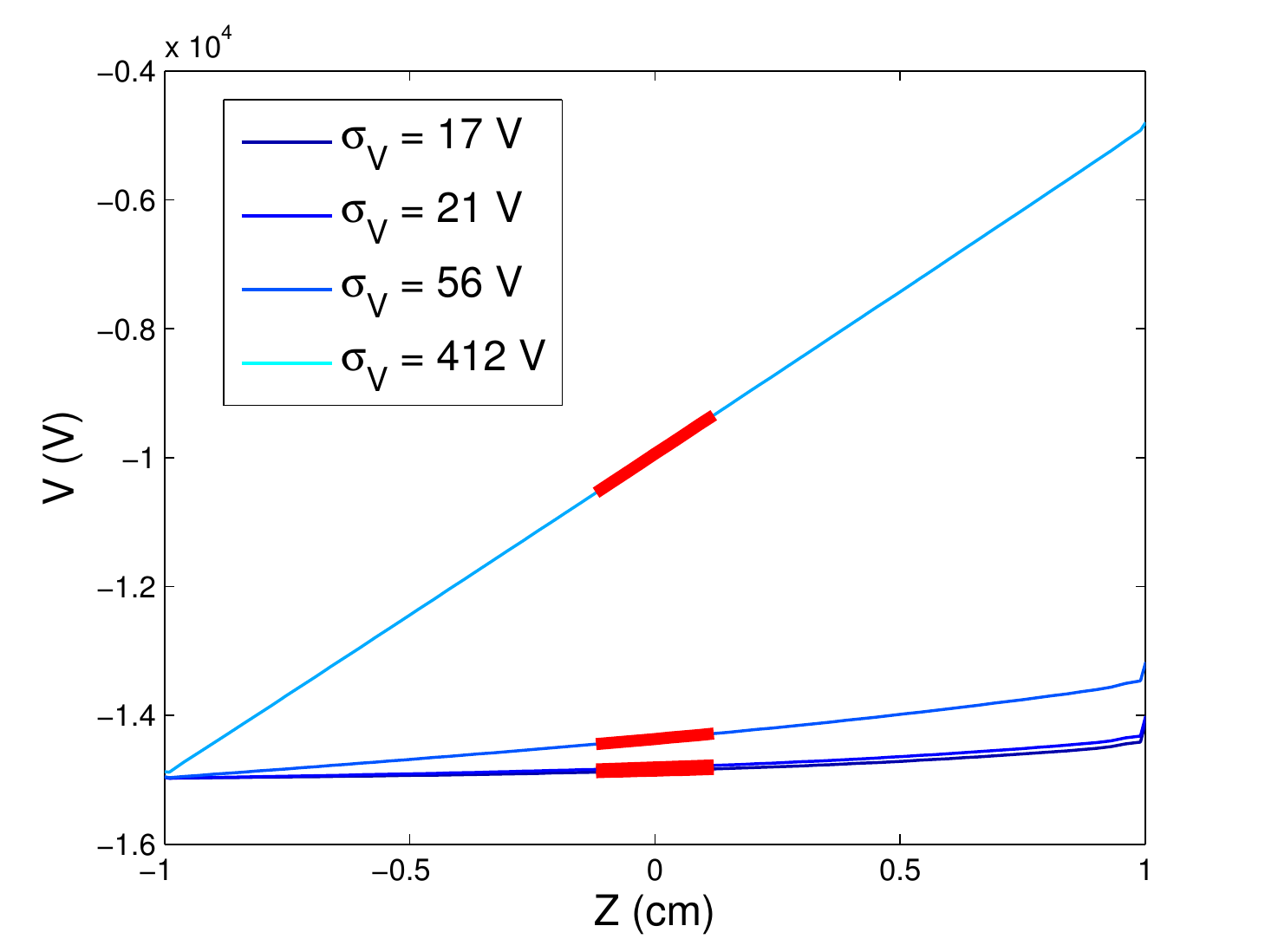}
\caption{Different voltages across the first two electrodes ($10\,V$, $100\,V$, $1\,kV$, $10\,kV$). The regions denoted by red correspond to gradients seen by the atom cloud.}
\label{fig:voltage_scheme1}
\vspace{-1.5em}
\end{figure}

Space charge induced beam emittance can be compensated by applying higher voltages on the electrodes without having to extract electrons over a large time scale. In order to accelerate the emerging electrons, electrodes are positioned and biased in order to provide a $5\,kV/cm$ and $15\,kV/cm$ gradients along the beam path in the first and second gaps separated by the electrodes, respectively. In the simulations all electrons are considered to be created at the same time, with a $1\,mm$ standard deviation around the centre of the MOT. 1000 macro particles corresponding to $1\,pC$ are tracked and their trajectories are presented in Fig.\ref{fig:traj1}. In Fig.\ref{fig:emitt_scheme1}, space charge dominated emittance is shown for various voltages between the first and the second electrodes as a function of position. Emittance scales down to $0.2\,mm\,mrad$ for the voltage configuration given in Fig.\ref{fig:traj1}, where there is $10\,kV$ voltage difference between the first two electrodes. The second and third electrodes are positioned close to each other to provide high electric field to rapidly accelerate and extract the electrodes out of the MOT. Nonetheless, the low emittance configuration compromises the initial energy spread of the beam. Fig.\ref{fig:voltage_scheme1} presents the standard deviation of the voltage across the atom cloud with a width of $\pm1\,mm$, for some voltage values corresponding to those in Fig.\ref{fig:emitt_scheme1}. Voltage deviations given are representing the standard deviation across the MOT.
\section{Conclusions and Outlook}
Pre-commissioning simulation results and operation modes of the cold atom trap based electron source in Manchester were presented in this paper. It has been, numerically, shown that it is possible to produce and extract electrons either with energy spread of $43\,meV$ or emittance of $0.2\,mm\,mrad$, initially, before beam optimisation to improve the beam quality even further. 

Forthcoming beam commissioning of the facility will initially assess beam profiles and intensities extracted from the system. The ionisation of the MOT with different laser wavelengths will be tested. The voltage difference across the MOT might be compensated by providing a large enough laser bandwidth rather than a monochromatic beam. This would allow creating a beam low energy spread and low emittance, simultaneously. 

Within the mid-term plans the facility will be modified for more sophisticated beam manipulation by including an RF acceleration section, focusing elements and a pepper pot emittance measurement station.  

The screening effect of positive ions in the MOT, present briefly after the ionisation of the electrons, will be numerically studied as it might mitigate a fraction of the space charge induced emittance growth. Momentum directions of the electrons can be initiated by the electric field vector of the laser ionising them. This phenomenon is also under numerical study alongside with the evolution of the different initial spacial distributions such as a hollow beam.

\end{document}